\documentclass[11pt,a4paper]{llncs}
\usepackage{epsfig} 
\begin{document}
\title{ Virtualization: A double-edged sword } 
\titlerunning{ Virtualization }
\author{ Joachim J. W{\l}odarz }
\authorrunning{ W{\l}odarz }
\institute{ 
Faculty of Mathematics, Physics and Chemistry, Silesian University,
Bankowa 14, 40-007 Katowice, Poland. 
Email: \email{ jjw@us.edu.pl } }
\maketitle
\begin{abstract} 
Virtualization became recently a hot topic once again, after being dormant
for more than twenty years. In the meantime, it has been almost forgotten, 
that virtual machines are not so perfect isolating environments as it seems, 
when looking at the principles. These lessons were already learnt earlier 
when the first virtualized systems have been exposed to real life usage.  
Contemporary virtualization software enables instant creation and destruction
of virtual machines on a host, live migration from one host to another,
execution history manipulation, etc. These features are very useful in
practice, but also causing headaches among security specialists, especially
in current hostile network environments.
In the present contribution we discuss the principles, potential benefits 
and risks of virtualization in a {\em d{\'e}j{\`a} vu} perspective, related
to previous experiences with virtualization in the mainframe era. 

\end{abstract}

\section{ Introduction }

Recently, a vigorous  interest in various virtualization techniques has been 
observed. Rather than being only a buzzword, virtualization became since 2005 
a main direction in the evolution of the whole IT market. According to 
recent predictions by the Gartner Group, operating system virtualization and 
I/O virtualization will become mainstream by 2010 \cite{GGP}.

Traditionally, the leadership in providing virtualization based solutions 
belonged to IBM, beginning with the VM/370 operating system for System/370 
mainframes introduced in 1972 \cite{IBMJRD25_483}. 
With the advent of relatively cheap microprocessor based computers within 
the next decade, the IT industry shifted their interests to personal computing,
based on individual hardware computers rather than some virtual machines.
But at the beginning of the 21st century, most personal workstations and
low-end servers became powerful enough to carry multiple operating systems 
running concurrently on the same hardware, which rised the interests 
in virtualization once again, this time on the commodity x86 platform. 
Moreover, it turned out from experience that setting up an extensive IT 
infrastructure with a lot of dedicated hardware put together with the 
appropriate, also dedicated software, is highly inefficient and also very 
expensive in management. Virtualization, leading generally to decoupling of 
software from the underlying hardware, seems to have the potential for 
substantial improvements in such situations.

In the meantime, virtualization software and virtualized systems became 
already a target of new security threats, with possibly disastrous consequences.
In contrary to the widespread belief that, e.g., virtual machines should be way 
more secure than their physical equivalents, virtualized computer systems may 
be sometimes {\em less} secure and even {\em create} new security challenges.

In the present contribution we discuss the benefits as well as the risks and
problems of employing virtualization based IT solutions. 
The discussion is undoubtedly biased due to the authors' experiences with
virtualization, first on various IBM mainframes many years ago and more recently
also on the x86 platform, where many problems appeared as {\em d{\'e}j{\`a} vu}.
As already emphasized in the title, virtualization is like a double-edged 
sword: a powerful weapon, but also quite dangerous for the warrior.

\section{Background}

When browsing through the literature, one could find several definitions for
virtualization, depending on the particular subject and its context. 
Generally, virtualization creates new entities in a computer system which 
are {\em substitutes} for the real ones, with the same functionality, 
interfaces and behavior, except the timings which are usually different.
The virtual resources, created through such a substitution process, have to 
be eventually mapped to appropriate real resources, possibly going through 
more than one level of indirection, if necessary.

The first and also the most important concept of a virtual entity in computer 
systems was virtual memory, introduced and developed around 1956 by 
Fritz-Rudolf G{\"u}ntsch \cite{ISp19_216} to unify various kinds of memory 
devices within an experimental computer system built at TU Berlin.
A couple of years later the first commercial computer with memory 
virtualization onboard was already built: the famous Atlas Computer 
\cite{MIT_FCHA_Atlas}
developed at the University of Manchester with collaboration of Ferranti/ICL. 
Although influential, the Atlas computer was known not to be working well in 
practical environments, mostly due to limited performance of the hardware
available at this time.

A more systematic approach to virtualization has been taken by IBM engineers 
working on the System/360, the first computer family "architected" according 
to common, precisely defined principles of operation. 
In order to develop an interactive, multi-user system based on the already 
existing S/360 operating systems, facilitate time-sharing and at the same 
time also protect the users from each other, they decided to provide each user 
with a dedicated "pseudo-machine", a fake System/360 computer capable to run 
any other S/360 operating system. 
The term "pseudo-machine" has been renamed soon to "virtual machine", when it 
turned out that the latter term has been already used earlier at IBM Research 
for an M44/44x experimental system, aimed at providing partially virtualized
``7044-like'' virtual machines on a modified IBM 7044 36-bit mainframe 
\cite{IBMSJ11_99}.

The intended full hardware virtualization per time-sharing user was quite
demanding and needed an appropriate support from the hardware itself to run 
with acceptable performance. 
Such hardware support appeared in only one System/360 computer: the famous
System/360 Model 67.
Regular hardware support for virtualization became reality a couple of years 
later, within the System/370 family and not without exercising some pressure 
from the ``VM community'' inside and outside IBM \cite{Melinda_V}. 
The first officially released version of VM/370 was able to provide an exact
functional copy of the underlying System/370 hardware, according to the
virtual machine configuration defined in the system directory. A particular
user could even run another copy of VM/370 on the virtual hardware, e.g., to
test a new version of system software without disrupting the running system.

The architecture of a ``classical'' VM/370 system, nowadays known rather as 
{\em hypervisor Type I architecture} according to a widely accepted 
classification, introduced by Robert P. Goldberg in his PhD thesis 
\cite{Goldberg}, is sketched in Fig.\ref{vm370} below. 
\begin{figure}[h] 
\begin{center} 
\epsfig{figure=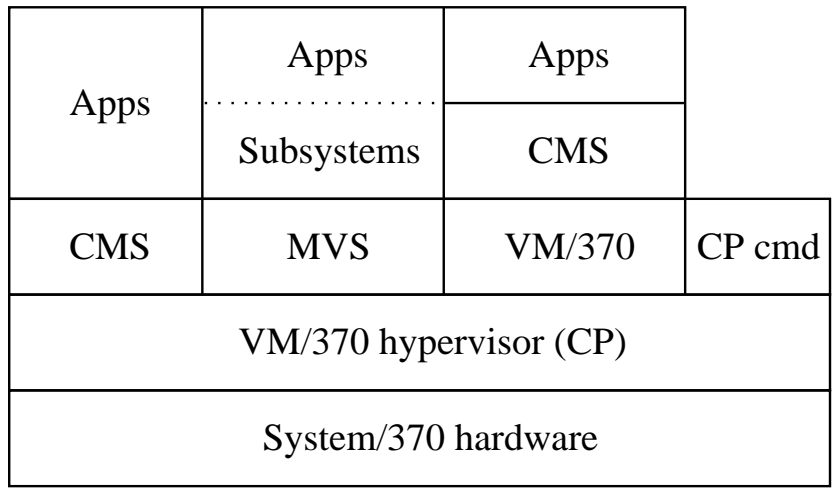} 
\end{center} 
\caption{The ``classical'' VM/370 hypervisor (Type I)} 
\label{vm370} 
\end{figure} 
The VM/370 hypervisor, or Control Program (CP) in the original IBM 
terminology, represents here the software layer implementing the virtual 
machines, each having exactly the same instruction set architecture (ISA) 
as the underlying System/370 hardware. 
This approach enabled a peaceful coexistence of several operating
systems with different needs and applications, running in parallel on the 
same hardware machine.
In the example presented in Fig.\ref{vm370}, one virtual machine runs CMS,
a simple operating system intended for individual usage, side-by-side with 
the mammoth-size MVS, a general purpose multitasking and multiuser operating 
system and even a second-level VM/370, running concurrently with its own 
virtual machines. 
A direct communication with the hypervisor was possible, and sometimes also
necessary, through the built-in CP command interface set up for each virtual
machine according to the assigned priviledge level.
The seemingly perfectly isolated virtual machines, running under the VM/370
hypervisor with carefully assigned priviledges, proved in practice to be 
totally insecure for intended intrusions, enabling even a total system 
penetration from a running VM, with direct access to the real hardware
and in the real supervisor state \cite{IBMSJ15_102}.

Virtual computer systems could be also created via emulation, where appropriate
software imitates another computer system or its part, e.g. CPU. In principle,
any operating environment could be emulated within another operating 
environment, including exact copy of the hardware used by the operating
system executing the emulator software itself. 
Emulators were originally intended to permit the execution of programs written
for another computer \cite{HoneywellCompJ6_287}, i.e. for another ISA than
employed on the host computer system, but there is certainly nothing wrong
with emulation of the same ISA and the related I/O hardware, giving
effectively another kind of hypervisor: the so called {\em Type II hypervisor}
\cite{Goldberg}.

\begin{figure}[h] 
\begin{center} 
\epsfig{figure=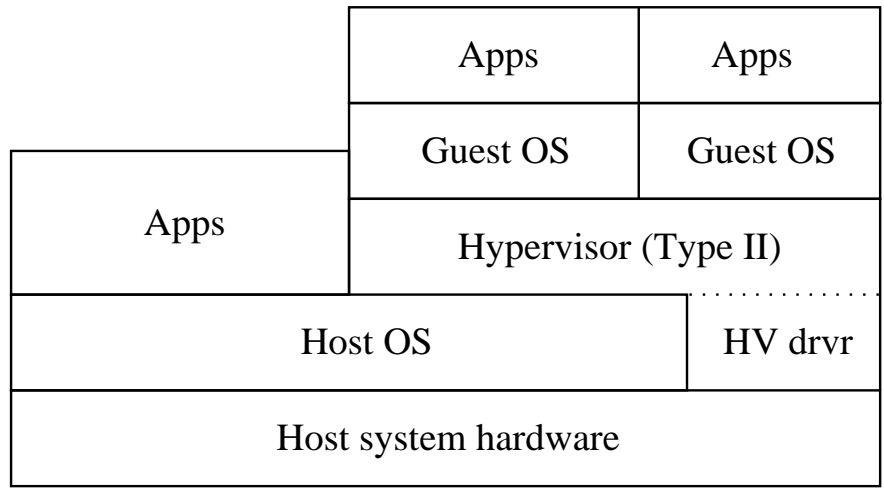} 
\end{center} 
\caption{A Type II hypervisor} 
\label{hv-t2} 
\end{figure}

Emulators and Type II hypervisors may run as ordinary applications within
the host computer operating system, being totally decoupled from the real
hardware. Although it may be desirable e.g. for debugging purposes, it means
usually a very slow execution within the virtual machine. 
A considerable speedup could be provided by the help of real hardware, made
available in parallel to the operating system services, usually through 
a specialized ``Hybrid Virtualization'' host operating system driver, see 
the sketch in Fig.\ref{hv-t2}.

\begin{figure}[h] 
\begin{center} 
\epsfig{figure=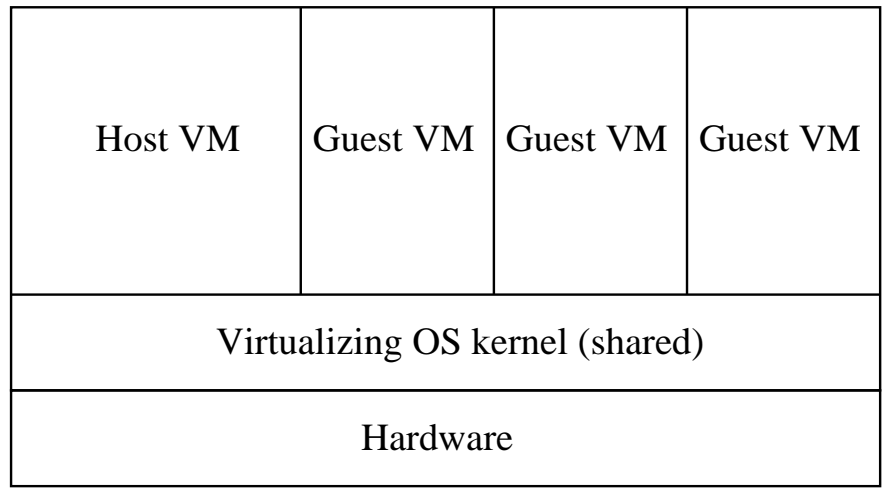} 
\end{center} 
\caption{Container-based operating system level virtualization } 
\label{cvm} 
\end{figure}

Yet another approach to virtualization, a so called 
{\em container-based virtualization},
is based on the mechanisms employed by contemporary operating systems
to establish a stardard execution environment for processes: 
the ``process virtual machine''.
Augmented with mechanisms for hiding and/or rewriting system-level data, 
the operating system kernel itself could be used as a kind of ``hypervisor'' 
to confine the processes within well-separated ``execution containers''. 
As shown in Fig.\ref{cvm}, the virtualized system contains in this case 
a fully priviledged ``Host VM'' used mainly for system management and several 
``Guest VMs'', which are seemingly equivalent to separate hosts.
The main advantage of container-based virtualization is high performance and 
scalability with only a minimal overhead introduced through the virtualization.
Many contemporary operating systems offer this possibility as a built-in or 
addon feature, e.g., Zones/BrandZ in Solaris, Secure Resource Partitions in
HP-UX or VServer and OpenVZ/Virtuozzo for Linux.

\section{x86 Virtualization}

The constantly increasing performance of x86-based computers within the
last decade rised also the interest in virtualization on this platform. 
The Intel Pentium ``classical'' IA-32 architecture has beed designed 
rather to remain compatible with previous Intel processors than being
virtualizable, with exception of the ``virtual 86'' sub-mode intended
for running ``Multiple Virtual DOS Machines'' in protected mode.
Therefore, it is quite challenging to implement virtualization software 
on this platform (cf. \cite{RobinIrvine2000,OpSysRev40_2} and references
therein).

Due to the recent high demand for virtualization, the newest Intel and AMD 
processors include a series of architectural extensions providing hardware 
support for full virtualization, which simplifies also considerably the 
implementation of x86-based ``true'' hypervisors (cf.\cite{ITJ10i3,AMD_SVM}).

The most hardware independent way of virtualization is pure emulation, where
no instructions are being passed directly to the host hardware. 
Several emulators exists for the x86 platform, the most advanced: 
Bochs \cite{Bochs} and QEMU \cite{QEMU} are available
for many host platforms, including x86 itself. QEMU enhanced with kqemu module
(a HV driver) works efectively as a Type II hypervisor with almost native speed.
The Kernel-based Virtual Machine, available with recent Linux kernels,  
provides full virtualization support with QEMU user-level frontend 
\cite{KVMwp}.

VMware Inc. \cite{VMware}, a pioneer of virtualization on the x86 platform, 
offers since 1999 a whole line of proprietary x86 virtualization software. 
The lowest-level products for workstations (VMware Player) and servers 
(VMware Server, earlier known as GSX Server) are available as freeware for 
MS Windows and also Linux. 
Similar virtualization products are offered also by Microsoft: Virtual PC and
Virtual Server, which became also freely available last year as direct
competitors to VMware products on the MS Windows platform.

The Xen hypervisor, developed initially as research project at the University
of Cambridge \cite{Xen2003}, seems to be the most promising and dynamically
evolving open-source virtualization software. Working generally as a Type I
hypervisor, Xen provides {\em paravirtualized} hardware for its virtual 
machines when full virtualization is not possible, e.g. when working on a 
Pentium host.

Paravirtualization enables a cooperation between the guest operating system and
the hypervisor to circumvent virtualization problems and increase performance.
The operating systems intended for the paravirtualized Xen virtual machines 
have to be ported to the Xen architecture, which may become problematic in case
of proprietary operating systems.
On the other hand, paravirtualized operating systems could perform even better
than their native versions running directly on the host hardware, because
the hypervisor may e.g. warrant apropriate resource allocations 
\cite{OpSysRev40_8}.

Virtual machines established under the Xen 3 hypervisor may be ``live migrated''
to another Xen 3 host computer, with only a very short application-level 
downtime around 100-300 ms in a local network with available network storage 
\cite{Xen2005}. 
It has been demonstrated recently, that a long-haul live migration is fairly
possible even at a planetary level, with downtimes rising only to 1-2 seconds 
for the distance between Amsterdam and San Diego \cite{FGCS22_901}.

\section{Applications and Benefits}

Virtualization technologies can be used for many purposes. In the mainframe
era the virtualized resources were localized on a single host computer and
accesible only to a relatively small number of users. 
Today, the possibility of migration of virtual machines from host to host, 
even in a ``live'' state, makes a tremendous difference. A virtual machine
could be instantly created on some host, then moved elsewhere, e.g., when 
the host becomes overloaded, or completely destroyed if not needed.
With a proper configuration, virtualized environments exhibit a sound level
of isolation from each other, hence a crash occuring in one environment
should remain unnoticed outside, which is a desired property for testbeds 
and security systems.

\subsection{Hardware Replacement}

Virtual hardware may easily replace real hardware consuming precious 
datacenter resouces like floor space, electricity or manpower. This
is especially important for sites running older equipment, which may 
be not easy to maintain, and at some point become even impossible to run,
e.g., due to the lack of appropriate spare parts on the market.
Virtualization is often used for server consolidation, i.e. a reduction 
of the total number of servers or even separate datacenters to a required 
minimum. 
At the same time, software which became incompatible with the new hardware 
and/or its configuration could be deployed within specialized virtual machines 
when needed, e.g., for archival purposes.
The possibility to migrate ``live machines'' across networks gives additional
flexibility in load balancing and disaster recovery, enabling  seamless
movement of workloads between distant datacenters  practically
without downtimes.

\subsection{Testing and Debugging}

Many virtualization systems, particularily emulators, were invented for
testing and debugging purposes, especially in the operating system kernel 
development, where the experimental system have to be rebooted again and
again. 
Also, getting some data for debugging may be not easy on a real hardware,
requiring sometimes arcane equipment. On emulated hardware, the same task
could be a trivial recording to some log file. With appropriate builtin
hooks, any virtualization environment may serve as a powerful debugger,
with possibility of a ``time travel'', i.e. arbitrary navigation in the 
execution history \cite{King05_1}.

\subsection{Education and E-Learning}

Virtual hardware is a very attractive possibility to have handy in any 
educational computer laboratory, because the students may experiment
with various operating systems and network setups, without fighting with 
problems related to running a multiplatform laboratory based on real hardware
\cite{SIGSE38_102}. 
For e-Learning purposes, such a ``virtual laboratory'' could be made
accesible remotely, e.g. from a standard web browser \cite{FOSLET06damiani}.

\subsection{Security Systems}

The isolation of applications running within a standard operating system is 
usually based only of the process abstraction, with many facilities available
for data sharing, also in a system-wide manner. 
Virtual machine technology may be therefore used to additionally isolate 
sensitive environments and minimize the risk of compromising the entire 
system due to problems in one of the environments.

Network services are usually regarded as particularily sensitive to security
problems, related to the direct contact with the possible hostile network
surrounding. 
Therefore, a separate network frontend virtual machine could be added to a 
virtualized system to handle the network connections and redirect them 
appropriately. With intelligent network hardware dedicated to this virtual 
machine and a minimal hardened operating system with integrated firewall, 
e.g. OpenBSD with pf, the overhead should be negigible \cite{Prevelakis}.

Much more demanding would be a virtual secure file server, established to
enforce a directory and file access control policy, even if the main system 
become compromised \cite{ZBP05b}.
It has been recently demonstrated that a more elaborated configuration, with 
several dedicated virtual machines and checkpointing, allows for an almost 
instant, automatical recovery after intrusion detection \cite{CNIS05_170}.

\section{The Other Edge of Virtualization}

Even in the mainframe era virtual machines were considered not secure enough 
for hosting sensitive data, mainly due to the I/O handling in VM/370 
\cite{IBMSJ15_102}. It turned out that a ``hardening'' by closing known
security holes does not guarantee the absence of other security flaws in the
hardened system, therefore a ``security retrofit program'' has been started
to design a formally verified ``security kernel'' for VM/370 \cite{vm370srp}.

There is no doubt that many hypervisors proliferating nowadays on the x86
platform need much more than only closing the detected security holes on the
fly. Without similar, specific ``security retrofit'' programs it would be 
pretty hard to avoid security problems. 
As shown in a very recent study by Tavis Ormandy \cite{CSW07ormandy},  
the present state of security exposure from implementation flaws in popular 
virtualization software is quite alarming. With exception of Xen, all tested
packages contained exploitable security flaws, allowing a potential attacker 
to escape reliably from the confining virtual machine.

The mobility of virtual machines, which could be easily migrated along multiple 
domains, adds its own vulnerabilities. E.g., a virtual machine may be moved 
into a faked recipient host (``kidnapped'') or compromised while stored 
semewhere in transit. 
The ease of manipulations involving virtual machines, which may be created
on the fly, possibly immitating another virtual/physical machine, then 
(mis)used for some purpose, and subsequently destroyed or moved away could 
be also a source of many security problems, unknown when working on dedicated 
hardware computers.
Static security arrangements are rather useless in this context.

As demonstrated recently \cite{King06}, it is perfectly possible to hide 
malware in a virtual machine running in parallel or force the host operating 
system into a virtual machine, keeping the malware running outside.
With latest hardware enhancements for virtualization, it is also possible
to do that on the fly, without disrupting the running operating system
\cite{BH2006JR,BH2006DDZ}.

\section{Conclusions}

When properly prepared and properly done, virtualization offers many
benefits in comparison with equivalent, contemporary available hardware 
based solutions. In the next few years virtualization will become a
standard way of computing in the enterprise, academia and personal use of
computers due to substantial savings and enormous flexibility.

The installation of operating systems and applications will quickly become
obsolete due to the possibility of preparation and distribution of virtual
appliances, tailored to the needs and ready to run on any computer with
the right hypervisor onboard. 

When using virtualization for security, a security enhanced implementation
with multilevel security \cite{MLS} would be definitely  recommended. 
A secure hypervisor architecture, sHype developed at IBM \cite{sHype}, 
targeted at the Common Criteria EAL4 assurance level, is now a part of the 
Xen open-source hypervisor \cite{XenMAC}. 
There are plans \cite{NRLVMM} for further elevation of the Xen assurance 
level at EAL5 and then EAL6. 
A prototype system for enforcing mandatory access control policies across 
a distributed set of virtual machines has been also tested recently with 
the Xen hypervisor \cite{Shamon}. 
Therefore Xen seems to be now the most advanced hypervisor with a proven
track record and clear evolution path in virtualization security. 

Last but not least, virtual machines are in principle nothing more than big 
fat programs with a complex inner life,  executed on a usually thin 
hypervisor layer, which in turn is in principle nothing more than a 
specialized operating system. Looking from this perspective, there is
no doubt, that virtualization software have to be always kept up-to-date 
and possibly equipped with a proper security add-on, exactly like ordinary 
operating systems. And the virtual appliances/machines have to be treated 
like ordinary executables, which are potentially corrupted, infected or 
malicious.

\end{document}